\documentclass[aip,twocolumn,showpacs,preprintnumbers,amsmath,amssymb]{revtex4}

\usepackage{graphicx}
\usepackage{dcolumn}
\usepackage{bm}

\begin{document}

\preprint{APS/123-QED}

\title{Optical tweezer for probing erythrocyte membrane deformability}

\author{Manas Khan, Harsh Soni and A.K. Sood$^{1}$}

 \affiliation{Department of Physics, Indian
Institute of Science, Bangalore - 560012, India}
\altaffiliation{$^{1}$ For correspondence : asood@physics.iisc.ernet.in}

\date{\today}

\begin{abstract}

We report that the average rotation speed of optically trapped crenated erythrocytes is direct signature of their membrane deformability. When placed in hypertonic buffer, discocytic erythrocytes are subjected to crenation. The deformation of cells brings in chirality and asymmetry in shape that make them rotate under the scattering force of a linearly polarized optical trap. A change in the deformability of the erythrocytes, due to any internal or environmental factor, affects the rotation speed of the trapped crenated cells. Here we show how the increment in erythrocyte membrane rigidity with adsorption of $Ca^{++}$ ions can be exhibited through this approach.

\end{abstract}


 \maketitle

In mammals, erythrocytes (red blood cells, (RBCs)) play a very critical role in transportation of oxygen and nutrition via microcirculation. In this process they are pumped through the capillaries of much smaller cross-sections. Thus any change in the deformability of the red blood cells affects the microcirculatory functions and may cause serious problems. The altered mechanical properties of the erythrocyte membranes often indicate the pathogenesis of many diseases \cite{pathogenic}. Sickle cell disease causes the RBCs to be less deformable and more fragile \cite{sickle}. Malaria infected red blood cells too show notable changes in the mechanical properties of the membranes \cite{pathogenic,malaria}. Hence, the estimation of the erythrocyte membrane deformability is of paramount importance. Many single-cell measurement techniques e.g. micropipette aspiration \cite{micropipette}, forced or flow induced stretching of an RBC with the help of optical tweezers \cite{flow,ss2003,radiation,nanomech,plasticflow}, microrheology study using optical magnetic twisting cytometry \cite{ss2007} and high-frequency electrical deformation tests \cite{ef} etc. have been used for this purpose. Rheological measurements also have been done on red blood cells suspensions \cite{drug,hemolysis,lectin,drochon}. Rotations of folded RBCs under circularly polarized optical trap have been observed too \cite{mathur}. Here, we report a simpler method to probe the erythrocyte membrane deformability, connecting it to the rotation speed of crenated RBCs under the scattering force of an optical trap.  

Normal healthy bi-concave shaped (discocytic) erythrocytes possess circular symmetry. While in an optical trap a discocytic erythrocyte stands up and aligns its diameter along the optic axis. Because of its symmetry, the trapped RBC does not experience any torque under the radiation pressure of the trapping laser beam and therefore it does not show any rotation. However when the cells are treated with hypertonic buffer, due to rapid efflux of water they get deformed, losing the inherent symmetry in their shape (crenation) \cite{crenation}. In most of the cases the asymmetric deformation brings in a finite chirality. A non-zero chirality about the optic axis converts the scattering force into a finite torque that makes the trapped body rotate about the axis \cite{galajda,nanorotor1,nanorotor2}. Thus most of the crenated erythrocytes show rotation in an optical trap \cite{pramana,mohanty}. The average speed of rotation depends on the chirality which, in general, increases with the contortion of the cells. 

With the tonicity of the medium, the degree of deformation of the RBCs changes and so does the average rotation speed \cite{mohanty}. On the other hand, if the intrinsic membrane deformability of the erythrocytes is changed by any internal or external condition, they would deform to a different degree albeit being in a similar hypertonic buffer. In this letter we report that the average rotation speed of crenated RBCs in an optical trap can be used as a measure of their membrane deformability. It is known that the presence of calcium ions $(Ca^{++})$ enhances the lipid ordering in erythrocyte membranes making them more rigid \cite{ca1,ca2,ca}. We have probed the same phenomenon through the average rotation speed and rotation probability of optically trapped crenated RBCs in hypertonic buffer containing different concentrations of calcium ions.

Fresh blood was obtained from healthy donors. The erythrocytes were separated by centrifugation and washed with phosphate buffer saline (PBS). Until the experiments, the cells were suspended in PBS (isotonic) and preserved at $4^{o}C$. Experiments were done within 36 hours after collecting the fresh blood samples. To treat the red blood cells with hypertonic buffer of osmolarity $1200$ $mOsm$, sodium chloride $(NaCl)$ and calcium chloride $(CaCl_{2})$ were mixed in different proportions and added to the original erythrocyte suspension in PBS ($300$ $mOsm$). In this method we obtained 5 different red cell suspensions in hypertonic buffer mediums of same osmolarity $1200$ $mOsm$ but with varying calcium ion concentrations in them. The relative proportions of $NaCl$ and $CaCl_{2}$ in the five different salt mixtures were determined in such a fashion that in the final suspensions the contribution of the $Ca^{++}$ ions to the total osmolarity of $1200$ $mOsm$ varied from $0\%$ to $25\%$, where the absolute value of $Ca^{++}$ ion concentration changed from $0$ to $300$ $mM$. To rule out the time dependence of the calcium ion treatment on the erythrocytes, each suspension was prepared and kept for one hour before it was loaded in the sample cell.

A linearly polarized $1064$ $nm$ $Nd:YVO_{4}$ diode pumped solid state laser was used to trap the erythrocytes. To built the optical trap, the laser beam was coupled to a modified Axiovert 200 microscope where a 63X special IR objective focused the beam tightly to the sample plane. A monochrome digital CCD camera, with $30fps$ capture rate, attached at the binocular port of the microscope along with an image acquisition card was utilized to visualize and record the observations. All the erythrocyte suspensions in different media were observed under the microscope before switching on the trap. Then the cells in hypertonic buffer solutions having different concentrations of $Ca^{++}$ ions were trapped at varying laser powers. To avoid the surface effect the trap was kept at 25-30 micron above the bottom plate. A total of 75 RBCs from each suspension were trapped at 5 different laser powers for better statistics. All the observations were recorded and stored in computer HDD for off-line analysis. For each of the suspensions, we counted the number of cells out of those 75 trapped RBCs that performed rotations under the trap. The rotation speeds were measured to get their average values as a function of trapping laser power for all the samples.

\begin{figure}[htbp]
\includegraphics[width=0.3\textwidth]{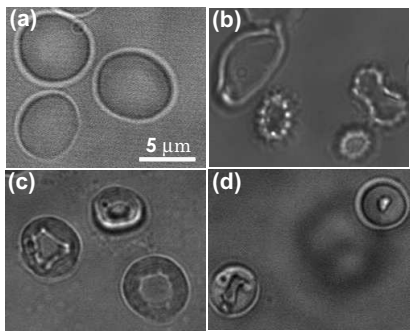}
  \caption{\label{rbc} The optical microscope images display erythrocytes in different conditions. (a) Normal healthy discocytic RBCs in phosphate buffer saline $(300$ $ mOsm)$. (b) Red blood cells in hypertonic buffer of osmolarity $1200$ $ mOsm$, without calcium ions. (c) RBCs in hypertonic buffer medium of same osmolarity $(1200$ $mOsm)$ but containing $ 150$ $mM$ of $Ca^{++}$ ions. (d) Erythrocytes in hypertonic buffer $(1200$ $mOsm)$ consisting of $300$ $mM$ of $Ca^{++}$ ions.}
\end{figure}

In the isotonic phosphate buffer, before adding the salt mixtures to make it hypertonic, the cells were biconcave disk shaped (discocytic) - as expected (Fig. \ref{rbc}(a)). The erythrocytes in the hypertonic buffer of $1200$ $ mOsm$ - without calcium ions, got deformed to a great extent. The crenated cells lost their disk like symmetric shape and became fairly contorted. With irregular morphology they almost attained the state of acanthocytes (Fig. \ref{rbc}(b)). The other hypertonic media (all were $1200$ $ mOsm$) with calcium ions did not cause that much of distortion to the symmetric shapes of the erythrocytes. As fluid came out of the RBCs, they shrank but retained their symmetry to some degree. The smaller and thinner cells looked more like a disk. The contortions were lesser for the erythrocytes in media containing higher concentrations of $Ca^{++}$ ions (Fig. \ref{rbc}(c),(d)).

\begin{figure}[htbp]
   \includegraphics[width=0.45\textwidth]{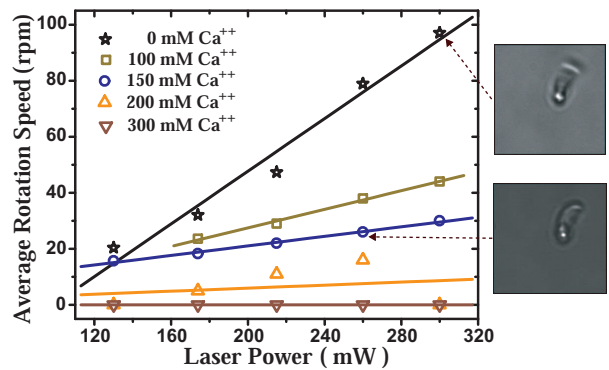}
  \caption{\label{speed} (Color Online) The average rotation speeds of the trapped crenated erythrocytes have been plotted against the laser power for five different calcium ion concentrations. For each concentration, the data points are fitted to a straight line. The open symbols represent the experimental data points and the solid lines are the linear fit to them. Typical rotations of trapped RBCs corresponding to two data points are shown in the side panel (enhanced online).}
\end{figure}

When trapped, most of the crenated cells started rotating. The RBCs with more deformation rotated faster. As each of the erythrocyte suspensions contained cells with a finite distribution in their sizes, ages etc, all the cells in the sample cell did not behave in the same fashion. Their degree of contortions and the speed of rotations showed a finite spread. Even some of them did not rotate at all though the others from the same sample performed rotation at moderate speeds under the trap. From the hypertonic buffer suspension of RBCs without any $Ca^{++}$ ions, $70\%$ of the trapped cells showed rotations. This percentage decreased as the concentration of the calcium ions in the suspension increased. Not a single erythrocyte rotated under the trap where the $Ca^{++}$ ion concentration was the most, $300$ $mM$. The number fraction of the total trapped crenated cells that performed rotation and their average rotation speeds are presented here (Fig. \ref{speed}) as a measure of the erythrocyte membrane deformabilities at different calcium ion concentrations.

\begin{figure}[htbp]
\includegraphics[width=0.35\textwidth]{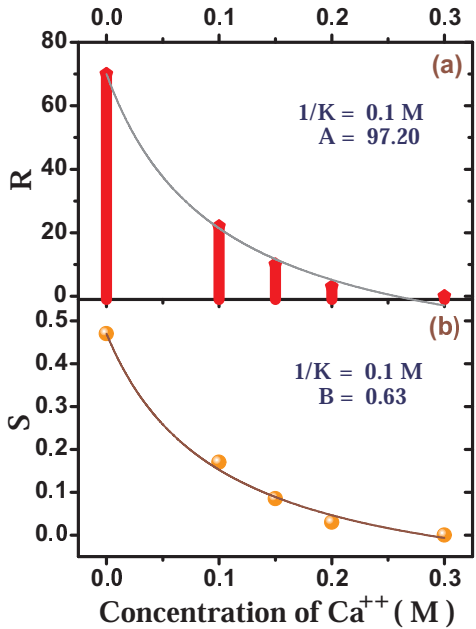}
\caption{\label{fit} (Color Online) (a) The bar graph shows $R$, the percentage of the total number of crenated erythrocytes that rotated under the trap, at varying $Ca^{++}$ ion concentrations, $C$. (b) Slopes ($S$) of the straight line fits in Fig. \ref{speed} are plotted against $C$. The dashed curves represent the fit to the data sets $R$ and $S$ following eqn 1a and 1b respectively.}
\end{figure}

For each of the suspensions, the average rotation speed of the trapped rotors increased linearly with the laser power, as expected for the asymmetric chiral rotors \cite{nanorotor1,nanorotor2}. (Two representative rotation observations are shown in Fig. \ref{speed} side panel.) Thus we get five straight lines for the five hypertonic media containing different concentrations of calcium ions (Fig. \ref{speed}). The slopes of the straight lines decreased monotonically with increasing $Ca^{++}$ ion concentration, becoming zero for the highest calcium ion concentration. The slope ($S$) that reflects the intrinsic membrane deformability of the RBCs, has been plotted against the calcium ion concentrations (Fig. \ref{fit}(b)). The other parameter that too reflects the deformations of the erythrocytes is the percentage ($R$) of the total number of trapped cells which performed rotation. The percentage, $R$, decreased with the increasing concentration of the calcium ion (Fig. \ref{fit}(a)), following the same trend as that of the former.

As the change in deformability of the erythrocytes with calcium ion concentrations is an effect of the adsorption of the ions to the cell membrane, the variation is expected to follow the Langmuir isotherm \cite{ca}. The isotherm relates the adsorption of molecules or ions ($\theta$, fractional coverage of the surface) on a surface to the concentration of the ions ($C$) as $\theta = K \cdot C/(1 + K \cdot C)$, where $K$ is the Langmuir adsorption constant. In our experiments, the observables - reflecting the erythrocyte deformability, drop off with increasing adsorption of the calcium ions. Therefore, to fit to our experimental data, we use two complementary equations:

\begin{subequations}
\begin{minipage}{0.45\linewidth}
\begin{equation}
R = R_{max}-A \cdot \theta 
\end{equation}
\end{minipage}
\begin{minipage}{0.45\linewidth}
\begin{equation}
S = S_{max}-B \cdot \theta
\end{equation}
\end{minipage}
\end{subequations}

$A$ and $B$ being fitting parameters. Fig. \ref{fit} displays the data sets $R$ and $S$ at varying $Ca^{++}$ ion concentration ($C$) and the corresponding fitting curves as solid lines. The parameters' values are given in the figure.

To conclude, the fitting of our experimental data shows that the variation of the average rotation speed of the trapped crenated RBCs and the probability of their rotation under the trap are the direct signatures of the red blood cell membrane deformability. The membrane deformability of the cells determine the degree of their contortions when subjected to hypertonic media of same osmolarity, and the contortions become visible through the rotation speed as well as the probability of rotation under an optical trap. Albeit being indirect, this method provides an easier alternative way to investigate the change of erythrocyte membrane rigidity and stiffness due to any internal or external abnormalities. Therefore, it could be used as a generalized diagnostic tool for the diseases that cause to affect the normal red blood cell membrane deformability.                  

We thank Department of Science and Technology, India and Council for Scientific and Industrial Research, India for financial support.

\end{document}